\documentclass[acmsmall,screen, nonacm]{acmart}

\usepackage{graphics}      
\usepackage[T1]{fontenc}   
\usepackage{booktabs}
\usepackage{textcomp}
\usepackage{enumitem}
\usepackage{dirtytalk}
\usepackage{multirow}
\usepackage{booktabs}
\renewcommand{\textrightarrow}{$\rightarrow$}

\setcopyright{rightsretained}
\copyrightyear{2022}
\acmYear{2022}

\begin{document}

\title[You, Me, and IoT]{You, Me, and IoT: How Internet-Connected Consumer Devices Affect Interpersonal Relationships}

\author{Noah Apthorpe}
\affiliation{
  \institution{Colgate University}
  \city{Hamilton}
  \state{NY}
  \country{USA}}
\email{napthorpe@colgate.edu}

\author{Pardis Emami-Naeini}
\affiliation{
  \institution{University of Washington}
  \city{Seattle}
  \state{WA}
  \country{USA}}
\email{pardis@cs.washington.edu}

\author{Arunesh Mathur}
\affiliation{
  \institution{Princeton University}
  \city{Princeton}
  \state{NJ}
  \country{USA}}
\email{amathur@cs.princeton.edu}

\author{Marshini Chetty}
\affiliation{
  \institution{University of Chicago}
  \city{Chicago}
  \state{IL}
  \country{USA}}
\email{marshini@uchicago.edu}

\author{Nick Feamster}
\affiliation{
  \institution{University of Chicago}
  \city{Chicago}
  \state{IL}
  \country{USA}}
\email{feamster@uchicago.edu}

\begin{abstract}
Internet-connected consumer devices have rapidly increased in popularity; however, relatively little is known about how these technologies are affecting interpersonal relationships in multi-occupant households. In this study, we conduct 13 semi-structured interviews and survey 508 individuals from a variety of backgrounds to discover and categorize how consumer IoT devices are affecting interpersonal relationships in the United States. We highlight several themes, providing exploratory data about the pervasiveness of interpersonal costs and benefits of consumer IoT devices. These results inform follow-up studies and design priorities for future IoT technologies to amplify positive and reduce negative interpersonal effects.
\end{abstract}



\maketitle
\renewcommand{\shortauthors}{N. Apthorpe et al.}

\section{Introduction} 
\label{sec:intro}

Consumer IoT devices have greatly increased in popularity over recent years and are often designed to replace existing non-networked products by introducing new effort-saving features into consumer homes.
Like the introduction of refrigerators, televisions, and other domestic technologies in
previous decades~\cite{OBrien}, the growing adoption of consumer IoT devices can dramatically
alter the day-to-day interactions between people living in shared spaces. 
Recent reports have documented that
IoT devices are disrupting households in unexpected ways---from replacing a spouse as an attentive conversation partner~\cite{Chelser} to being used by domestic abusers to exert control over others in their homes~\cite{Bowles, Havron}. 

In this study, we investigate how consumer IoT devices affect interpersonal
relationships, including how they improve household
dynamics and how they cause or exacerbate interpersonal conflicts. 
We use the terms ``Internet of things'' and ``IoT devices'' in this paper to refer to consumer-grade Internet-connected physical products designed predominantly for domestic use, excluding smartphones, tablets, personal computers, and Internet-connected technologies in non-commercial domains (e.g., industrial, commercial, or medical). This aligns with previous definitions of the consumer Internet of things~\cite{bitag} and encompasses the broad variety of devices considered as such in the literature, including voice assistants~\cite{He}, game consoles~\cite{Mazhar}, smart TVs~\cite{Son}, WiFi speakers~\cite{Hatzivasilis}, security cameras~\cite{Apthorpe-keeping}, large appliances~\cite{Li}, activity trackers~\cite{Jalal}, and other ``smart home'' automation devices~\cite{Copos}.
This inclusive definition allows us to consider a wide range of IoT devices that intersect with many aspects of users' lives. However, we do not assume that the devices owned by our study participants are comprehensive of the consumer IoT space.
We use the term ``interpersonal relationships'' in this paper to refer to social interactions, connections, and opinions existing over an extended period among multiple individuals sharing a household or other physical space. We use the terms ``interpersonal benefits'' and ``interpersonal conflicts'' to refer to events and actions that strengthen or weaken these relationships, respectively. These definitions align with the vernacular use of these terms and are consistent with ideas expressed in prior research on shared IoT technology use~\cite{Geeng, Zeng, Kraemer}.

We first conducted 13 semi-structured one-on-one interviews with individuals living in multi-occupant U.S.~households with a variety of IoT devices  (Section~\ref{sec:interview-method}). 
The interviews involved discussions of how IoT devices have affected household relationships from a variety of perspectives, including spouse/partner/roommate dynamics, parenting decisions, and interactions with guests. 
Open-coding of interview transcripts revealed several recurring themes that
deepen our understanding of IoT devices and interpersonal relationships.
We then surveyed an additional 508 individuals 
living in multi-occupant households with IoT devices to better understand
the extent of the effects surfaced in the interviews and to identify additional themes
across a larger sample size and wider variety of demographics (Section~\ref{sec:survey-method}). 

The combined interview and survey results indicate that IoT devices often \textit{benefit~(B)} interpersonal relationships and cause interpersonal \textit{conflict~(C)} by the following mechanisms (Section~\ref{sec:results}):

\begin{itemize}
\item[\textbf{B1.}] \textbf{Strengthening interpersonal connections} through bonding over shared experiences, simplifying remote communication, and inspiring playfulness. 
\item[\textbf{B2.}] \textbf{Enabling empowerment and independence} by reducing the sense of being a burden and helping individuals with special needs. 
\item[\textbf{B3.}] \textbf{Easing household management}, resulting in increased free time to spend with household members and improved peace of mind. 
\item[\textbf{C1.}] \textbf{Facilitating surveillance and causing mistrust} due to potential or actual undesired monitoring and a lack of data collection transparency.
\item[\textbf{C2.}] \textbf{Provoking differences in knowledge or preferences} about the functionality, benefits, risks, privacy, or security of IoT devices.
\item[\textbf{C3.}] \textbf{Causing tensions about device use, sharing, and technical issues} that arise during day-to-day operation and proximity of the devices. 
\end{itemize}

These results are important, because qualitative research on IoT devices and household relationships remains limited, and large-scale quantitative data about the interpersonal effects of consumer IoT adoption are likewise non-existent in the HCI literature (Section~\ref{sec:related}).
Revealing and categorizing these interpersonal conflicts and benefits allows us to identify common underlying factors that not only motivate future studies but also inform recommendations for device manufacturers 
(Section~\ref{sec:discussion}).
First, insufficient and unclear documentation leads to users having contradictory mental models of device behaviors and conflicting assumptions about data collection practices and appropriate device use.
Second, many IoT devices lack customization options with enough flexibility to account for diverse user relationships, especially in households where individuals have different device control responsibilities or data privacy concerns.
Manufacturers must be cognizant of these
issues
while recognizing
that IoT devices, when designed well, can actually improve
interpersonal relationships. 
By enhancing device documentation, clarifying data collection practices, and providing more flexible customization options, manufacturers could better support real-world use of their products.
Ultimately, this paper forms the basis for future 
investigations of the interpersonal benefits and conflicts we report and serves as a call for manufacturers to consider a broader range of social and household dynamics when designing IoT devices.

This paper makes the following contributions:
\begin{itemize}
    \item Discovers and categorizes common effects of IoT devices on interpersonal relationships through interviews and open-ended survey responses.
    \item Provides exploratory survey data indicating the pervasiveness of  interpersonal conflicts and benefits across multi-occupant U.S. households.
    \item Discusses common underlying factors, recommendations for device manufacturers, and follow-up studies to improve the
    effects of IoT devices on interpersonal relationships.
\end{itemize}
\section{Related Work}
\label{sec:related}

Most research to date has only tangentially examined how consumer IoT devices affect interpersonal relationships, often in light of related research topics, such as multi-user content sharing or privacy concerns.
A few closely related studies conducted in 2019~\cite{Geeng,Kraemer,Zeng} investigated multi-user interactions and shared
control of IoT devices in homes. 
Other IoT user studies have focused on different research questions,
including purchasing decisions~\cite{Emami2019}, privacy concerns
regarding entities external to the home (manufacturers, governments,
etc.)~\cite{Zheng}, privacy expectations of devices
themselves~\cite{Apthorpe-discovering,Emami2017}, and how friends
and experts influence IoT data collection consent~\cite{Emami2018}.
Our project complements and extends this literature by 
specifically focusing on the interpersonal benefits provided by IoT devices as well as the household tensions, conflicts, or
disagreements caused by these products. 

\subsection{Benefits of IoT Devices}
Previous studies of the benefits of IoT
devices have focused predominantly on functionality with fewer studies noting how these devices benefit interpersonal relationships.

\subsubsection{Curiosity \& Routines}
Lazar et al.~\cite{Lazar} found that interview participants chose to use IoT devices because ``the devices satisfied curiosity and held hope for
potential benefit to them,'' or because the devices had been incorporated into
the participants' routines. Our work also indicates that curiosity about home
IoT devices can improve interpersonal relationships by inspiring playful behavior and facilitating bonding over shared interests (Section~\ref{sec:interpersonal-connections}).

\subsubsection{Convenience}
Coskun et al.~\cite{Coskun} found that improved comfort and performance 
through automation incentivized the incorporation of IoT appliances into households.
Zheng et al.~\cite{Zheng} also found that early adopters cited
convenience as a primary reason for using IoT devices, a factor that
outweighed concerns about privacy vis-a-vis device manufacturers, governments,
and other entities external to the home. 
Strengers, et al.~\cite{Strengers} similarly noted that productivity benefits were central to
experiences with IoT devices for 31 early adopters, including small conveniences such as energy savings and support for multi-tasking. 
This paper extends these findings by showing that the conveniences afforded by IoT devices can directly benefit interpersonal relationships (Section~\ref{sec:easing-tasks}).

\subsubsection{Connection with Friends and Family}
Emami-Naeini et al.~\cite{Emami2019} found that prospective buyers of IoT devices often turned to friends and family for word-of-mouth recommendations and advice. 
Woo and Lim~\cite{Woo} conducted an observational study in DIY smart homes and found that home automation could provide emotional comfort as a happy reminder of the person who set up the automation. 
Takayama et al.~\cite{Takayama} found that home automation systems can support family communication, connection to loved ones, and positive household monitoring (e.g., observing a family pet when away from home). 
Strengers et al.~\cite{Strengers} reported that early adopters appreciated IoT device features that allowed them to better protect their households, viewing this protection as a form of care provided to others in their home. These early adopters also cited improvements to home ambiance provided by IoT devices and the ability to showcase new technologies to visitors. Kraemer et al.~\cite{Kraemer} described the processes used by a household navigating shared IoT devices as ``group efficacy,'' extending Bandura's definition of self-efficacy~\cite{Bandura} to communal behavior. 
Morris~\cite{Morris} provides many examples of individuals using connected technologies to support and enhance social relationships, often in ways not anticipated by the technology designers. While some of these studies, especially~\cite{Morris}, prioritize varying \textit{uses} of technology, others (including this paper) explore the effects that connected devices have on relationships.

\subsubsection{Community Benefit}
An ethnographic study by Burrows et al.~\cite{Burrows} found that users of IoT health technologies were willing to share anonymized data if they believed it would improve community well-being. This corroborates findings by Zheng et al.~\cite{Zheng} that early adopters were willing to share some IoT data with local governments to improve utility expenses and other services for the entire community. While not the focus of this study, these findings indicate how IoT devices could positively affect interpersonal relationships outside of the household. 

\subsection{Conflicts Involving IoT Devices}
Existing research has also examined how IoT devices cause interpersonal conflicts, typically regarding specific use cases or topics of contention (e.g., privacy).

\subsubsection{Power Imbalance and Technical Expertise} 
Geeng and Roesner
studied shared control of IoT devices in different living
situations~\cite{Geeng} and found that multi-user tensions can be
categorized by when they occur, namely during ``(1) device selection and
installation, (2) regular device usage, (3) when things go wrong, and (4) over
the long-term.'' They also provide examples of tensions arising in specific
partnership, roommate, and parent/child relationships and note that many of
these tensions are caused by differences in ``power, agency, technical skill,
and technical interest.'' Some studies have also found that Internet-connected
products may amplify domestic disputes and abuse~\cite{Bowles,Freed,Havron}.  
Our work is consistent with these
results---we find similar concerns over surveillance, for example---and adds
further context to past work by exploring the prevalence of these concerns.
More generally, we focus on a broader set of interpersonal relationships
beyond control and power dynamics. We also provide new examples of
interpersonal conflicts involving IoT devices and quantitative data
indicating the pervasiveness of these and other causes of tensions (Sections~\ref{sec:monitoring}--\ref{sec:use-sharing}).

\subsubsection{Incompatible Incentives}
Zeng and Roesner~\cite{Zeng} conducted an interview study and
design exploration to understand how users navigate security and privacy
issues in multi-occupant homes with IoT devices. They found that users wanted access
controls in place for configuration changes, parental controls, and devices in
private rooms---all indicating situations in which different household members
may have differing incentives that could lead to conflict. They also note the
importance of social norms, trust, respect, lack of concern, and a desire for
convenience as inhibitors of access control use---factors that we find also provide interpersonal benefits in homes with IoT devices.

\subsubsection{Differences in Knowledge and Expectations}
In 2018, Malkin et al.~\cite{Malkin} found that there was a great deal of uncertainty and assumptions about smart TV data collection practices among surveyed users. 
In 2017, Zeng et al.'s~\cite{Zeng2017} interviews found that differences in security/privacy mental threat models, differences in access and control of IoT devices, and surveillance all led to disagreements or concerns in multi-user homes.
In 2014, Ur et al.~\cite{Ur} interviewed parents and children about their opinions of home-entryway surveillance and observed a disconnect between parents' and children's surveillance preferences, which could potentially cause interpersonal conflict.

In 2012, Mennicken and Huang~\cite{Mennicken} observed variations in roles, including ``home technology drivers,'' ``home technology responsibles,'' and ``passive users.'' Users in these categories had qualitatively different knowledge of and experience with home technologies. 
Our participants also had
a range of knowledge and preferences regarding IoT device behavior, 
complicating the categories of Mennicken and Huang~\cite{Mennicken} by demonstrating the diversity of household relationships and roles. These results provide further interpersonal relationship context to Brush et al.'s 2011 results on UI and access control from DIY smart homes~\cite{Brush} and show that some of these issues continue with mass-market IoT products.

\subsubsection{Changing Privacy Norms}
Issues of privacy in shared spaces often arise in studies of consumer IoT devices, including in many works cited above. Researchers have framed these issues using formal privacy theories, including the application of contextual integrity~\cite{nissenbaum2009privacy} to understand the landscape of sensitive data and privacy concerns in smart homes and smart buildings~\cite{mccreary2016contextual, Apthorpe-discovering} and quantified-self health data~\cite{patterson2013contextual}. The rapidly changing landscape of consumer IoT products is creating new privacy norms and expectations for shared spaces, a topic explored by Zafiroglu, et al. in 2016~\cite{zafiroglu2016living} and raised by several of the participants in this study (Section~\ref{sec:monitoring}).
\section{Interview Method}
\label{sec:interview-method}

We conducted 13 semi-structured interviews to understand how consumer IoT devices are affecting interpersonal relationships in multi-occupant households. The interviews involved a scripted series of questions interspersed with and followed by open-ended conversation. 

The interview study was approved by 
the Princeton University and Carnegie Mellon University
Institutional Review Boards (IRB). All participants provided their informed consent to participate in the screening survey and interviews, to have their voice recorded, and to have the recordings transcribed by a third-party company. We anonymized the transcriptions prior to coding.

\subsection{Recruitment}
We recruited participants through Craigslist in the 
Central New Jersey and Pittsburgh, Pennsylvania
regions containing our universities. We also used snowball recruiting, asking interviewees to recruit their friends, family, and acquaintances. The Craigslist advertisements stated that ``researchers at 
Princeton and Carnegie Mellon 
Universities want to better understand your interactions with smart (Internet-connected) devices'' and ``researchers at 
Princeton and Carnegie Mellon 
Universities want to better understand how smart (Internet-connected) home devices and appliances can cause disagreements, tension, or conflict in interpersonal relationships between people living in the same household.''
The advertisements specified that participants must be at least 18 years of age and live in a home or apartment with at least one other person and at least one IoT device. 

The advertisements invited individuals to complete a short screening survey. The screening survey asked respondents to list the number and relationships of people living in their household, the number and types of IoT devices in their household, and how they acquired those devices. It also included a series of demographics questions, including age, gender, income, education, occupation, and technology background.

The advertisements were online for five days, after which the screening survey responses were reviewed and qualifying respondents were contacted for interview scheduling. We received 77 responses from Craigslist recruiting. We also received 2 responses from snowball recruiting. 
We selected all 51 respondents who reported owning at least one IoT device and living with at least one other person. We emailed these respondents with two tentative dates and times for interviews that fit with their reported availability. 26 respondents replied to confirm an interview time. Of these, 13 participants actually joined the video call for the interview at the scheduled time, resulting in 13 total interviews. 

\begin{table*}[t]
\scriptsize
\centering
\caption{Interview participant demographics, household occupants, IoT devices, and interview durations (mm:ss).}
\begin{tabular}{llllllll}
\toprule
 & \textbf{Gender} & \textbf{Age} & \textbf{Income} & \textbf{Education} & \textbf{Occupants} & \textbf{IoT Devices} & \textbf{Duration}\\
\midrule
PI1 & M & 24 & $<$\$20k & College & 3 Roommates & 6 security cameras, smart TV & 17:54\\ 
PI2 & F & 42 & $>$\$100k & College & Domestic partner & Amazon Echo & 28:11\\
PI3 & M & 22 & $<$\$20k & High School & Domestic partner & Amazon Fire TV, gaming consoles & 19:08\\
PI4 & F & 41 & \$50-75k & College & Spouse, 2 Children & Amazon Echo, Amazon Echo Dot, & \\
& & & & & & \quad Google Home, Sonos & 21:17\\
PI5 & M & 50 & $>$\$100k & High School & Domestic partner & Amazon Echo & 21:46\\
PI6 & F & 22 & \$50-75k &  Prof. Deg. & 2 Roommates & Google Home & 18:57\\
PI7 & M & 58 & $>$\$100k & Assoc. Deg. & Spouse & Amazon Echo, TVs, Amazon Fire Stick, & 20:39\\
& & & & & & \quad refrigerator, washer, dryer, doorbell & \\
PI8 & F & 53 & \$50-75k & Prof. Deg. & 1 Child & Amazon Echo, security cameras, smart TV & 15:43\\
PI9 & F & 21 & $<$\$20k & College & 2 Roommates & Roku TV & 20:29\\
PI10 & F & 21 & $>$\$100k & High School & Domestic partner & Google Home, August Smart Lock & 19:39\\
PI11 & F & 30 & $>$\$100k & Prof. Deg. & Domestic partner & Amazon Echo, Amazon Show, smart TV & 16:07\\
PI12 & M & 36 & $>$\$100k & College & Spouse, 3 Children & Amazon Echo, Roku, wireless doorbell & 15:35\\
PI13 & F & 34 & \$50-75k & College & Spouse, 1 Child & Amazon Echo Dot, iRobot Roomba, & \\
& & & & & & \quad smart TV, smart plugs & 16:40\\
\bottomrule
\end{tabular}
\label{tab:participants}
\end{table*}

These 13 participants had a range of demographic backgrounds, living situations, and IoT devices in their households (Table~\ref{tab:participants}). There were 5 male and 8 female participants ranging from 22 to 58 years old. The participants lived with roommates, spouses, significant others, and children. They owned many popular IoT devices, including voice assistants, smart TVs, IoT locks, WiFi appliances, and others.  All participants were compensated with a \$25 Amazon gift card for completing the interview.

\begin{table}[t]
\small
\caption{Interview script. The interviewer asked the device-specific questions about one to three IoT devices in the participants' households as time allowed. The interviewer also asked participants to freely expand on topics when appropriate given the semi-structured nature of the interviews.}
\begin{tabular}{p{0.22\textwidth}p{0.72\textwidth}}
\toprule
\textbf{Category} & \textbf{Questions} \\
\midrule
Household & \begin{minipage}{0.7\textwidth}
\begin{itemize}[nosep, leftmargin=8pt]
\item Who lives in your household?
\item What Internet-connected devices do you have in your home?
\end{itemize}\end{minipage}\\
\midrule

Device Purchasing & 
\begin{minipage}{0.7\textwidth}
\begin{itemize}[nosep, leftmargin=8pt]
\item How long have you had the device? 
\item Who purchased the device and why? 
\item Did you have any concerns about the device before purchase?
\end{itemize}\end{minipage}\\
\midrule

Setup \& Accounts & 
\begin{minipage}{0.7\textwidth}
\begin{itemize}[nosep, leftmargin=8pt]
\item Who set up the device?
\item Who is in charge of managing the device?
\item Do you have individual or shared accounts on the device?
\end{itemize}\end{minipage}\\
\midrule

Device Use & 
\begin{minipage}{0.7\textwidth}
\begin{itemize}[nosep, leftmargin=8pt]
\item How and why do people in your household use the device? 
\item How well do you and others understand how to use the device?
\end{itemize}\end{minipage}\\
\midrule

Benefits & 
\begin{minipage}{0.7\textwidth}
\begin{itemize}[nosep, leftmargin=8pt]
\item Has the device improved the relationships between people in your household? If so, please describe.
\item How else has the device benefited people in your household?
\end{itemize}\end{minipage}\\
\midrule

Conflicts & 
\begin{minipage}{0.7\textwidth}
\begin{itemize}[nosep, leftmargin=8pt]
\item Has the device been involved in any conflicts, tensions, or disagreements in your household? If so, please describe.
\item Who in your household was involved in these conflicts? 
\item Were these existing conflicts or new ones caused by the device?
\item Did you take any steps to mediate these conflicts?
\end{itemize}\end{minipage}\\
\midrule

Privacy & 
\begin{minipage}{0.7\textwidth}
\begin{itemize}[nosep, leftmargin=8pt]
\item Have you discussed or disagreed about the privacy implications of the device with others in your household?
\item Have you ever used the device to monitor others?
\item Do you think others have ever used the device to monitor you?
\end{itemize}\end{minipage}\\
\midrule

Children \mbox{(if applicable)} & 
\begin{minipage}{0.7\textwidth}
\begin{itemize}[nosep, leftmargin=8pt]
\item Do your children use this device? 
\item Have your relationships with your children improved due to the device?
\item Have you had any conflicts, tensions, or disagreements with your children about their use of the device?
\end{itemize}\end{minipage}\\
\midrule

Design Feedback &
\begin{minipage}{0.7\textwidth}
\begin{itemize}[nosep, leftmargin=8pt]
\item What is your opinion about the device? 
\item What would you like to change about the device?
\end{itemize}\end{minipage}\\

\bottomrule
\end{tabular}
\label{tab:interview-script}
\end{table}

\subsection{Interview Procedure}
All interviews were conducted on a one-on-one basis by the first author over video call and were semi-structured in nature. 
The interviewer used a prepared script (Table~\ref{tab:interview-script}) and followed up on topics that arose naturally during the conversation, leading to discussions that varied widely depending on the opinions and experiences of each participant.
The interview script 
included questions about
household occupants and devices, device purchasing, setup and account management, device use by home occupants, 
interpersonal benefits involving the device,
interpersonal conflicts involving the device,
privacy and in-home surveillance, device use by children, and device design feedback.
When discussing interpersonal benefits and conflicts, the
interviewer guided the conversation to 
ensure that the participant reported which devices were involved, how household members were affected, whether the device contributed to existing conflicts or created new conflicts, and whether any steps were taken to mediate the conflicts. 
The interviews only focused on participants' relationships as appropriate. For example, participants without children were not asked about children's interactions with their devices. All interviews lasted between 15--30 minutes.

\subsection{Data Analysis}
\label{sec:interview-response-analysis}
We transcribed the interview audio recordings using NVivo's automated transcription service~\cite{nVivo} then
manually reviewed the transcriptions, making corrections as necessary to ensure accuracy. 
We performed open coding~\cite{Seidman} on the transcriptions to identify recurring themes. Two authors independently arrived at a set of codes and then consolidated their codes into a codebook\footnote{Interview codebook provided in the Supplementary Material} with 6 main parent codes: ``Positive experiences,'' ``Benefits to relationship,'' ``Conflicts \& concerns,'' ``Conflict mediation,''
``Involvement,'' and ``Time of benefit/conflict.'' We also had a total of 40 child codes. For example, the ``Time of benefit/conflict,'' parent code had child codes ``purchase time,'' 
``installation time,'' and ``use.'' Each interview transcript was coded by these two authors and disagreements were discussed and resolved in multiple meetings. The entire research team met regularly to identify the main concepts and themes occurring across the coded data. These themes informed the questions in the follow-up survey (Section~\ref{sec:likert-questions}) and are reported along with additional themes from the survey as the primary results of this study (Section~\ref{sec:results}). 
We did not calculate inter-rater reliability (IRR) for our interview analysis because the coded data was
not an end product but a process used to derive concepts as themes, making an IRR measure unnecessary in this case~\cite{McDonald_2019}.

\section{Survey Method}
\label{sec:survey-method}

We conducted a survey to measure the pervasiveness of the interpersonal effects of IoT devices observed in the interviews and to discover additional themes across a wider variety of demographics. 
The survey was approved by 
the Princeton University and Carnegie Mellon University 
Institutional Review Boards. All respondents
provided their informed consent to participate in the survey. 

\subsection{Survey Design}
The survey contained five sections:\footnote{Full survey provided in the Supplementary Material}

\subsubsection{Consent Form and Home Context}
The survey began with a consent form. Respondents were then asked to indicate the number of the people in their household, the relationships of these people to themselves (e.g., ``spouse'' or   ``parent''), and the types of IoT devices in their household.
Respondents selected their IoT devices from a multiple-choice list of the Internet-connected products in Table~\ref{tab:survey-demographics}. This list was provided by the custom prescreening options of the survey deployment platform (Section~\ref{sec:deployment}). 
This facilitated survey deployment and provided a broad view of IoT devices consistent with our definition in Section~\ref{sec:intro}.
All respondents who did not agree to the consent form, had no IoT devices, or lived alone were not allowed to continue the survey and were not included in the results analysis. 

\newcommand{\rowgroup}[1]{\hspace{-0.5em}#1}
\begin{table}[t]
\scriptsize
    \centering
        \caption{Self-reported demographics, living situations, and IoT devices of survey respondents. The less prevalent ``other devices'' include smart doorlocks/doorbells, baby cameras/monitors, smart water sprinklers/irrigation controllers, smart health monitors, smart smoke monitors and alarms, smart kitchen appliances, and smart Bluetooth trackers.}
    \begin{tabular}{rl}
        \toprule
         \textbf{Demographic} & \textbf{Sample}\\
         \midrule
         \rowgroup{\scriptsize{\textbf{Gender}}} \\
         Female & 53\% \\
         Male   & 46\% \\
         Other & 2\% \\
         \\
         \rowgroup{\scriptsize{\textbf{Age}}} \\
         18--24 & 19\% \\
         25--34 & 42\% \\
         35--44 & 21\% \\
         45--54 & 11\% \\
         55--64 & 6\% \\
         65--74 & 1\% \\
         75$+$ & $<$1\% \\
         \\
         \rowgroup{\scriptsize{\textbf{Education}}} \\
         No high school & 1\% \\
         High school & 34\% \\
         Associates & 11\% \\
         College & 39\% \\
         Prof. deg. & 14\% \\
         Prefer not to disclose & 1\% \\
         \bottomrule
         \end{tabular}%
    \begin{tabular}{rl}
    \toprule
         \textbf{Demographic} & \textbf{Sample}\\
         \midrule
         \rowgroup{\scriptsize{\textbf{Individual Annual Income}}} \\
         $<$\$20,000 & 9\% \\
         \$20,000--\$34,999 & 13\% \\
         \$35,000--\$49,999 & 17\% \\
         \$50,000--\$74,999 & 20\% \\
         \$75,000--\$99,999 & 18\% \\
         $>$\$100,000 & 20\% \\
         Prefer not to disclose & 3\% \\
         \\ 
         \rowgroup{\scriptsize{\textbf{Household Size}}} \\
         2 people & 39\% \\
         3 people & 24\% \\
         4 people & 23\% \\
         5 people & 8\% \\
         6$+$ people & 6\% \\
         \\
         \rowgroup{\scriptsize{\textbf{Language at Home}}} \\
         Only English & 86\% \\
         Other language & 13\% \\
	\\
	\\
         \bottomrule
         \end{tabular}%
    \begin{tabular}{rl}
    \toprule
         \textbf{Demographic} & \textbf{Sample}\\
         \midrule
         \rowgroup{\scriptsize{\textbf{Household Members}}} \\
         Spouse & 48\% \\
         Child & 36\% \\
         Parent & 24\% \\
         Partner & 16\% \\
         Other relative & 15\% \\
         Housemate or roommate & 9\% \\
         Other non-relative & 2\% \\
         \\
         \rowgroup{\scriptsize{\textbf{IoT Devices}}}\\
         Games console & 75\% \\
         Smart TV & 64\% \\
         Video streaming product & 60\% \\
         Home assistants/smart hub & 43\% \\
         Activity tracker & 33\% \\
         Smart watch & 21\% \\
         Connected lights & 14\% \\
         Smart security camera & 13\% \\
         Smart thermostat & 13\% \\
         Smart plugs & 11\% \\
         Other devices & <10\% \\
         \bottomrule
    \end{tabular}%
    \label{tab:survey-demographics}
\end{table}

\subsubsection{Interpersonal Relationship Questions}
The next section of the survey asked respondents whether ``Internet-connected products have caused any disagreements (major or minor) between people in your household?'' Respondents who answered ``yes'' were asked to describe the conflict in an open-ended text response question and then to answer multiple choice questions about which device(s) had been involved in the conflict, who in the household had been involved in the conflict, and what steps (if any) they had taken to mitigate the conflict. Respondents who answered ``no'' were asked to describe whether they ``have had any other negative experiences with Internet-connected products.''

This structure was then repeated for interpersonal benefits, first asking respondents whether ``Internet-connected products have improved your relationships with others in your household?'' Respondents who answered ``yes'' were asked to describe this improvement in an open-ended text response question and then to answer multiple choice questions about which device(s) and household members were involved in the improved relationship. Respondents who answered ``no''  were asked to describe whether they ``have had any other positive experiences with Internet-connected products.'' 

\subsubsection{Likert-scale IoT Questions}
\label{sec:likert-questions}
The following section contained a matrix of Likert-scale multiple choice questions with the prompt ``How much do you agree with the following statements about home technology?'' and five answer choices: ``Strongly agree,'' ``Somewhat agree,'' ``Neither agree nor disagree,'' ``Somewhat disagree,'' and ``Strongly disagree.''

The statements were generated from recurring themes in the interviews in order to measure their pervasiveness across a larger sample size. Examples include ``Internet-connected products have inspired playful behavior in my household'' and ``I have disagreed with others in my household about whether we should have Internet-connected products in our home.''
We used the interview participants' own wording about benefits and conflicts (e.g., ``disagreed'' and ``tensions'') when creating these survey questions to facilitate interpretability. These questions were not intended to be of balanced valence between positive and negative effects and should not be interpreted as such.
Figures~\ref{fig:bvc-likert}, \ref{fig:benefits-likert}, and \ref{fig:conflicts-likert} present the full list of statements with response distributions. 
This section also included an attention check question asking participants to select ``Somewhat disagree.''  
After viewing the Likert-scale questions, respondents could not return to modify their answers to the open-ended questions. This prevented priming effects from the Likert question prompts from influencing open-ended responses.

\subsubsection{Demographics}
The survey concluded with a series of standard demographics questions, including age, gender, education, 
annual household income, and primary language spoken at home. 

\subsection{Survey Deployment and Respondent Overview}
\label{sec:deployment}
We tested the length and clarity of the survey by performing seven 10-minute ``cognitive interviews'' on UserBob~\cite{userbob}, a usability testing platform that recruits crowdworkers at a rate of \$1/minute to interact with a website while recording their screen and providing audio feedback.
We asked the workers to ``go through the survey, pretending you are a participant and letting us know whether the survey makes sense.''
We adjusted the survey based on their feedback, including reducing the number of questions per page and adding bold font to highlight the Likert-scale questions.
The UserBob recordings confirmed that respondents interpreted the questions as expected, avoiding the need for wording changes.
The UserBob responses were not included in the final results.

We recruited 536 respondents through Prolific~\cite{prolific}, an online survey service founded in 2014 that maintains its own pool of respondents and emphasizes data quality, fair compensation, and significantly fewer bot-like accounts than Amazon Mechanical Turk~\cite{Bradley}. We chose Prolific because it allowed us to pre-screen for respondents with multi-occupant households and reported ownership of Internet-connected products. This prevented the need for a separate screening survey as would have been necessary on Amazon Mechanical Turk. 
All respondents were paid $\$1.10$ for completing the survey, resulting in an average compensation of $\$13.20$/hour across all respondents. 

The survey respondents all lived in the United States and had a variety of demographic backgrounds, living situations, and IoT devices (Table~\ref{tab:survey-demographics}). The respondents were 53\% female, 82\% younger than 45, 53\% with college degree or higher, 39\% with individual annual incomes less than \$50,000/year, and 61\% living in households with more than two individuals. 
This higher proportion of young, well-educated respondents compared to the general population reflects well-known biases in Internet use in the United States~\cite{pew-internet-use} and other crowdsourcing platforms~\cite{Hitlin}.  
The potential effects of these and other representativeness issues on the survey results are discussed in Section~\ref{sec:limitations}.

\subsection{Response Analysis}
\label{sec:survey-response-analysis}
We started with 536 survey responses. We removed 16 responses that failed the attention check question and 12 responses from those who took less than two minutes to complete the survey. The remaining 508 responses used for analysis had a median completion time of 5.85 minutes.

We performed open coding~\cite{Seidman} on the open-ended text responses. Two authors independently coded these questions, consolidated their codes into interpersonal benefits and conflicts codebooks (Tables~\ref{tab:codebook1}--\ref{tab:codebook2}), then re-coded the questions, achieving a Kupper-Hafner intercoder reliability score~\cite{Kupper}  greater than 0.76 on all questions for a sample of 100 respondents. We used these final codebooks to identify several of the interpersonal conflicts and benefits themes presented in Section~\ref{sec:results}. 

\begin{table*}[t]
\scriptsize
\centering
\caption{Codebook for open-ended responses to the survey questions ``Describe how Internet-connected products have improved your relationships with others in your household. Please provide as much detail as you can.'' and ``If you have had any other positive experiences with Internet-connected products, please describe them here.''}
\begin{tabular}{l l}
\toprule
\textbf{Code} & \textbf{Explanation} \\
\midrule
play & Playfulness and entertainment leading to bonding \\
convenience & Convenience and improving quality of life\\
connected & Staying connected with family and friends\\
do more & Ability to do more or having more choices and features\\
financial & Saving money together\\
time & Enabled spending time together (includes conversation, bonding, etc.)\\
health & Staying fit together\\
security & Enabling safety and security\\
special pop.~& Helpful for people with disabilities or special needs\\
interactions & Fewer interactions with each other leading to fewer conflicts \\
none & None\\
not clear & Not clear\\
\bottomrule
\end{tabular}
\label{tab:codebook1}
\end{table*}

\begin{table*}[t]
\scriptsize
\centering
\caption{Codebook for open-ended responses to the survey questions ``Describe how Internet-connected products have caused disagreements (major or minor) in your household. Please provide as much detail as you can.'' and ``If you have had any other negative experiences with Internet-connected products, please describe them here.''}
\begin{tabular}{l l}
\toprule
\textbf{Code} & \textbf{Explanation}\\
\midrule
choice & Hard to choose the right device (due to various specifications)\\
f2f & Negative effects for face-to-face communication\\
functionality & Functionality and technical challenges of setting up IoT devices\\
misbehavior & Misbehavior caused using IoT devices\\
necessity & Lack of need, interest or perceived benefit in technology or IoT devices \\
network & Discussions around bandwidth sharing\\
parenting & Challenges in parenting caused by kids' use of devices\\
privacy & Privacy and comfort related concerns (e.g., surveillance, data use, data sharing,\\
& \quad discomfort caused by shared privacy settings)\\
unexpected & Unexpected device behavior\\
updates & Difficulties caused due to firmware updates and troubleshooting\\
variance & Different set of users of the same device and their varying use preferences\\
none & None \\
not clear & Not clear\\
\bottomrule
\end{tabular}
\label{tab:codebook2}
\end{table*}

\begin{figure}[t]
    \centering
    \includegraphics[width=0.85\textwidth]{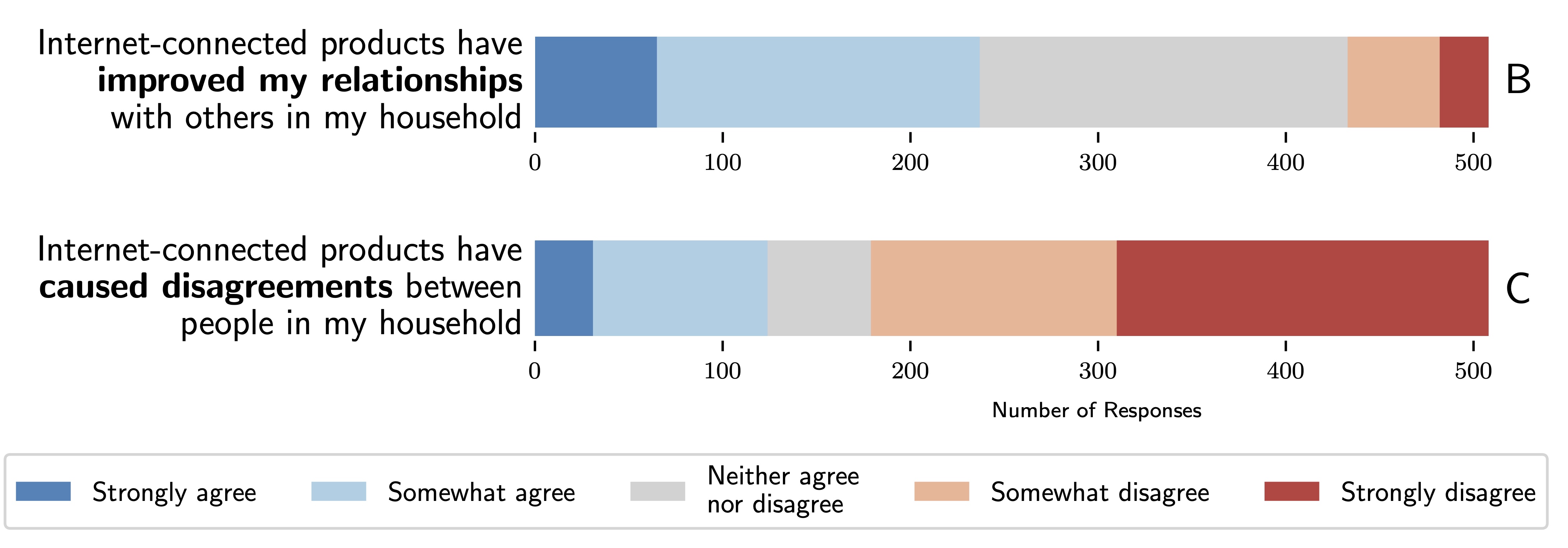}
    \caption{Survey responses indicating the prevalence of interpersonal benefits (B) and interpersonal conflicts (C) resulting from IoT devices.}
    \label{fig:bvc-likert}
\end{figure}

We then analyzed the multiple choice questions to determine the pervasiveness of these themes (Figures~\ref{fig:bvc-likert}--\ref{fig:disagreement-resolution}). We compared the relative prevalence of interpersonal benefits versus conflicts by applying the Chi-squared test to compare the distributions of responses to the questions ``Have Internet-connected products caused any disagreements (major or minor) between people in your household?'' and ``Have Internet-connected products improved your relationships with others in your household?''
We also compared the responses to selected Likert-scale multiple choice questions across demographic groups, using Mann-Whitney $U$ tests to compare the distribution of agree responses (``strongly agree'' or ``somewhat agree''), neutral responses, and disagree responses (``strongly disagree'' or ``somewhat disagree'') to each question of interest between all pairwise sets of respondents with different answers to each demographic question.

Given the small interview sample size, we did not compare results between the surveys and the interviews. Rather, we combined the qualitative and quantitative evidence provided by both methods into our results and discussion (Sections~\ref{sec:results}--\ref{sec:discussion}).
\section{Results}
\label{sec:results}

\begin{figure}[t]
    \centering
    \includegraphics[width=0.7\textwidth]{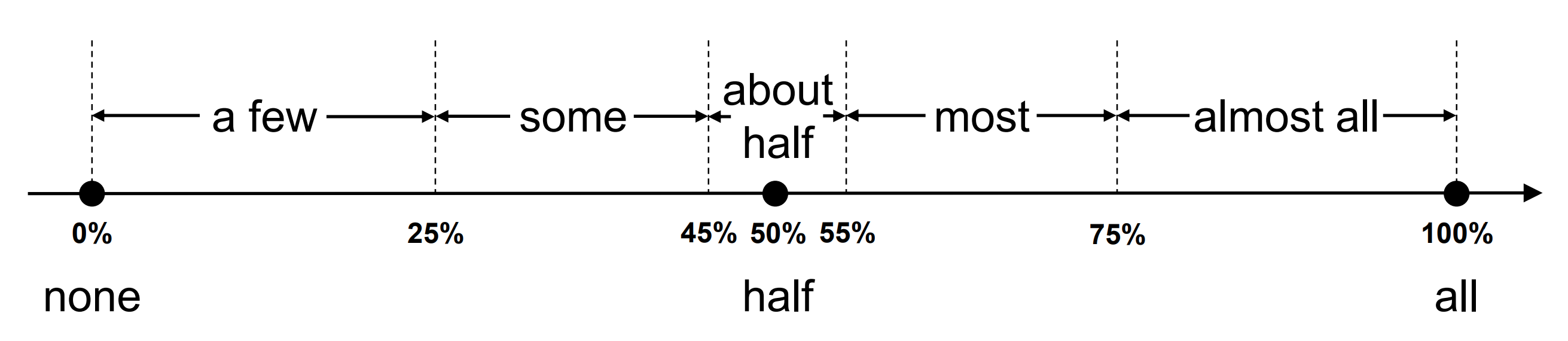}
    \caption{Qualitative terminology used to report findings of interviews and open-ended survey questions. Figure from~\cite{Emami2019}.}
    \label{fig:qualitative-terminology}
\end{figure}

Our interviews and survey responses indicate the richness of interpersonal benefits (B1--B3) and
conflicts (C1--C3) involving consumer IoT devices. This section provides quantitative and
qualitative data to support the pervasiveness and
influence of these themes and their importance to IoT adoption, design,
and research. 
We refer to interview participants as PI\#, survey respondents as PS\#, and use the qualitative terminology from Emami-Naeini et al.~\cite{Emami2019} to report the frequency of qualitative findings from the interviews and the open-ended survey questions (Figure~\ref{fig:qualitative-terminology}).
We also present data about conflict mediation and other ways that users are adapting their lives with IoT devices. 

\subsection{Interpersonal Benefits Versus Conflicts} 
Significantly more survey respondents reported that IoT devices have improved
their relationships with others in their household (49\%) compared to those who reported
that IoT devices have caused disagreements in their
household (23\%, $p \ll 0.01$). This corroborates the higher frequency of ``strongly agree'' and
``somewhat agree'' responses to the corresponding Likert-scale questions about
relationship improvements versus conflicts (Figure~\ref{fig:bvc-likert}). 
We did not find any significant differences between the reported frequency of interpersonal benefits or conflicts across demographics, indicating that while such variations may exist, a larger or more representative group of respondents would be necessary to identify them given their effect size. 

The interpersonal benefits reported by our participants range from the well-studied, such as simplifying
everyday tasks~\cite{Zheng}, to the less-understood, such as helping
support a household member with special needs.
Although the reported interpersonal conflicts are less frequent, they are often serious,
including the use of devices to surveil household members without their
knowledge: 9\% (46/508) of survey respondents report active disagreements with
others in their household about surveillance, and 15\% (78/508) agree that they
have used Internet-connected products to monitor someone else's behavior.

\begin{figure}[t]
    \centering
    \includegraphics[width=0.85\textwidth]{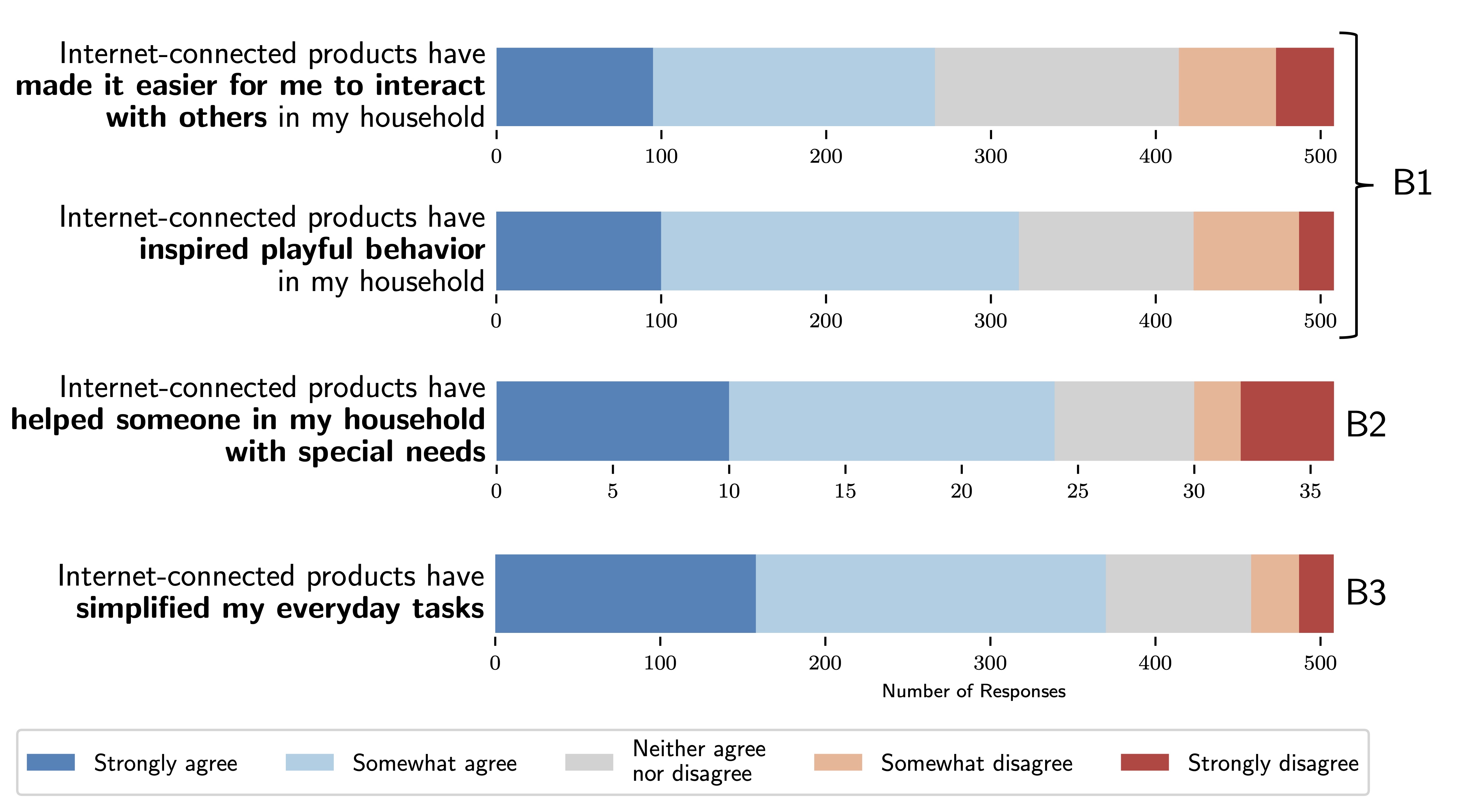}
    \caption{Survey responses indicating the prevalence of interpersonal benefits from IoT devices related to themes B1--B3. Note the difference in scale for the third question, which was only asked of respondents who reported having special needs individuals in their households.}
    \label{fig:benefits-likert}
\end{figure}

\begin{figure}[tp]
    \centering
    \includegraphics[width=0.88\textwidth]{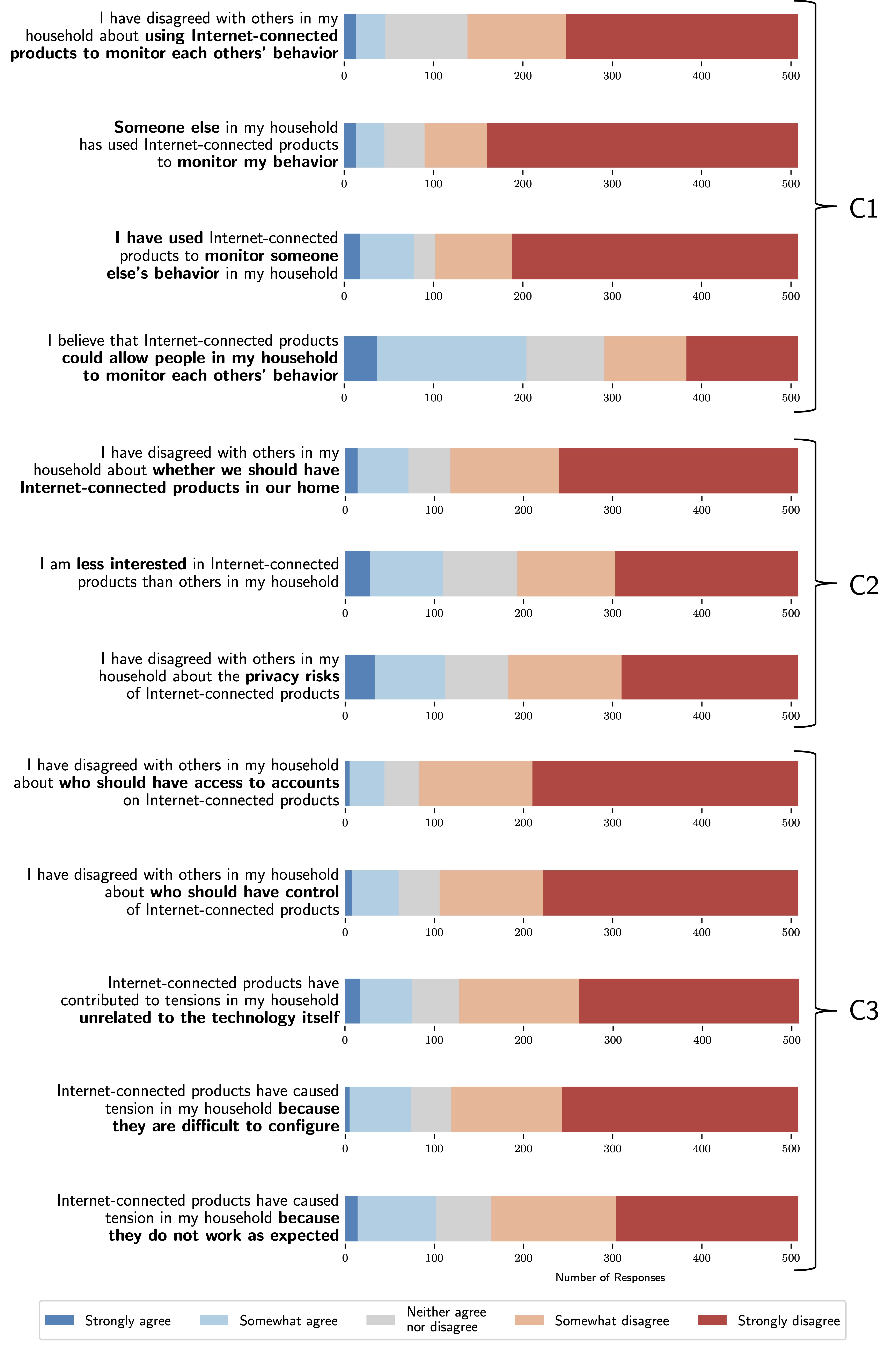}
    \caption{Survey responses indicating the prevalence of interpersonal conflicts involving IoT devices related to themes C1--C3. The questions are sorted within each theme by the number of ``agree'' responses.}
    \label{fig:conflicts-likert}
\end{figure}

\subsection{B1.~Strengthening Interpersonal Connections}
\label{sec:interpersonal-connections}
Most participants who reported positive experiences with their IoT
devices linked these experiences to improved interpersonal connections with other household members. We found several recurring ways that these devices facilitated such strengthened connections.

\subsubsection{Bonding Over Shared Experiences}
Most of our interview and survey participants said that IoT devices
caused family members to bond over shared experiences, 
often facilitated by the ease of content sharing enabled by the devices.
For example, 
PS97 said,
\begin{quote}
Streaming movies helps my relationship with my partner. It gives us bonding time.
\end{quote}
PS438 talked about similar positive
experiences with an IoT speaker: 
\begin{quote}
Smart devices made it easier to
share music with my siblings, like smart speakers for example. Instead of
having to pass someone's phone or rely on one person connected, we can just
tell it to play a song and boom.
\end{quote}
IoT devices also precipitated inter-generational bonding when a younger generation helped an older relative with technology they found too complicated, as PI2 described:
\begin{quote}
We've got an Apple TV and my father almost cried because he said he was really curious about [the device] and streaming television, but he felt too out of the loop and overwhelmed to try another giant leap in technology. And he was overjoyed...to have my boyfriend help out with setting it up.
\end{quote}
PS73 described helping relatives with IoT devices and bonding over this kind of support even more succinctly:
\begin{quote}
My parents are not exactly tech savvy, so when I help them in terms of the use of technology, it becomes a kind of bonding moment.
\end{quote}
More than 50\% (266/508) of survey respondents agreed that
Internet-connected products made it easier to interact with others in their
household (Figure~\ref{fig:benefits-likert}). IoT devices necessitate setup, management, and maintenance, and if these responsibilities are distributed amongst household members or family members in different houses, they can facilitate increased communication and connection. 

\subsubsection{Simplifying Remote Communication}

Some of our participants reported that their IoT devices helped them keep
in touch with their remote family members. PS381 described this benefit as provided by Amazon Echo and Google Home voice assistants:
\begin{quote}
I am better able to stay
connected to my adult children and to my disabled husband when I am at
work.  
\end{quote}
PI5 described a similar situation involving communication with his mother through an Amazon Echo instead of having her try to find and work her phone:
\begin{quote}
    My mother was sick\dots and before she passed away, it was tougher and tougher for her to use the phone\dots So what I did was I got an Alexa and I installed it in the house, and then I could just call her and rather than her having to figure out how to answer the phone, she could just hear my voice in the ether.
\end{quote}
This quote reveals a benefit provided by a consumer IoT device over a more traditional phone interface.
IoT voice assistants also helped a few survey respondents, including PS21, communicate ``remotely'' with family members inside their home:
\begin{quote}
Communicating with my kids is so much easier when we put Echo Dots on each level of our house. We can just drop in on each other and talk instead of yelling.
\end{quote}
These findings corroborate past research showing the positive impacts of
technology-mediated communications
between household members, such as by conveying messages
via changing color light bulbs~\cite{Morris}. In this case, IoT devices made interactions between family members in the same or different households easier because they mimicked more natural voice interactions.

\subsubsection{Inspiring Playfulness} 
A few of our interview participants noted that their IoT devices, particularly
voice assistants, inspired inquisitive and playful behavior among
members of their household. This playfulness was often expressed as asking non-serious
questions to the voice assistant to entertain others in the household. For
example, PI7 said that their Amazon Echo Dot,
\begin{quote}
Lets us sit around and laugh at the
different answers\dots almost like playing a game.
\end{quote}
Similarly, PI2 said that hearing their boyfriend talking to their Alexa was amusing:
\begin{quote}The main joy that I get from Alexa is overhearing my boyfriend ask her
ridiculous things just to see like if she'll respond, how she'll respond.
\end{quote}

These participants shared these anecdotes as some of their favorite experiences with IoT products. The playful feature exploration invited by these voice assistants was related to the perceived novelty of the voice interface.
Playfulness was also prevalent factor in positive interpersonal impacts of IoT devices for survey respondents. 62\% (317/508) reported that their Internet-connected products have inspired playful behavior
in their household (Figure~\ref{fig:benefits-likert}). 

\subsection{B2.~Enabling Empowerment and Independence}
\label{sec:independence}
Most of our participants reported that their IoT devices helped family members seek information and enhance their knowledge without relying on other household members. As PS129 reported, 
\begin{quote}
My wife can now just ask the Google Home for the weather instead of assuming I know what the weather is.
\end{quote}
Strengers et al.~\cite{Strengers} described how IoT technologies could help individuals ``live independently in their own homes for so many more years.'' Our results indicate that these benefits are not limited to those living alone, but that improved independence provided by IoT devices can also benefit interpersonal relationships in shared households.

For some of our participants, IoT devices, especially voice assistants, helped family members with special needs when looking for information on their own. PI4 described this benefit for their son who uses Alexa for answering questions:
\begin{quote}
    My youngest son  is actually autistic, but he's very inquisitive in nature and asks me the most intelligent but random questions that we can never really answer. So it's always like ``Go ask Alexa''\dots It's almost like having a  teacher or an encyclopedia like right on hand at all times, and for his way of living that's just really helpful for him.
\end{quote}
PS445 also described how streaming services accessed through a smart TV helped their child with special needs:
\begin{quote}
My kids are special needs, and the ability to find teaching videos through [smart TV] streaming apps has been incredibly valuable to helping teach basics as well as social skills.
\end{quote}
The potential benefits of IoT devices for households with special needs individuals was further corroborated by the multiple-choice survey responses. 66\% of the 36 survey respondents who reported an individual with special needs in their household also agreed that their Internet-connected products had helped that individual (Figure~\ref{fig:benefits-likert}).

\subsection{B3.~Easing Household Tasks}
\label{sec:easing-tasks}
Prior work has found that people would like their household chores to be
automated, as they perceived them as unwanted
tasks~\cite{Coskun, Eggen}. Most of our
participants reported that their IoT devices provided 
convenience in routine tasks and helped them achieve more efficient time
management in the household. This was especially predominant in the survey
responses: 73\% (370/508) of respondents agreed that their Internet-connected
products had simplified their everyday tasks
(Figure~\ref{fig:benefits-likert}). Convenience is a well-studied
individual benefit of IoT devices~\cite{Page, Zheng}. This study extends these
previous findings, demonstrating interpersonal benefits gained from improved convenience.

\subsubsection{Increasing Free Time with Household Members}

Most of the participants who reported convenience as one of the benefits of
IoT devices also said that this convenience allowed them
more time to spend with their family
members. When asked about the positive experiences of having these devices,
PS182 mentioned that their IoT device
\begin{quote}
Freed us up to be able to
spend more time catching up with each other.
\end{quote}
PS50 likewise said that IoT devices make a household easier to manage:
\begin{quote}
Having ``smart'' technology makes it easier to run and manage our household, giving us more time to focus on one another.
\end{quote}

\subsubsection{Reducing Tensions About Household Management}
Some of our participants noted that their IoT devices reduced arguments about chore responsibilities and day-to-day household management.
In some cases, these participants were able to entirely offload tasks to their IoT devices. 
PS325 described how allowing an IoT thermostat and doorbells to automatically manage parts of the home environment reduced household tension:
\begin{quote}
With the smart thermostat, we don't argue about the temp of the house because it's automatically set...With the doorbells, we don't have to argue or wonder if it was locked. We can just look on the app...All the small conveniences add up to a happier and healthier lifestyle so we have less tension in the household over stuff.
\end{quote}
PS231 described nearly identical benefits of delegating tasks to IoT devices instead of relying on household members to do these tasks:
\begin{quote}
We don't have to nag each other to get up and do something. We can ask the device to do it for us. We are not getting into arguments on who forgot what and who didn't set the temperature or lock the door. Everything is programmed.
\end{quote}
In other cases, IoT devices helped household members keep track of day-to-day tasks, preventing the need for other members to remind them.
This benefit was typically attributed to IoT voice assistants, as described by PS332:
\begin{quote}
My partner and I use Amazon Echo to set reminders for each other, which helps with making sure we are both on the same page with groceries and chores.
\end{quote}
PS341 also described how automated reminders improved their relationship with their children:
\begin{quote}
I have the Amazon Echos in my kids' rooms set to remind them to do daily things like get ready for bed and straighten their rooms. By not having to personally nag them to do these things, we get along better on a daily basis.
\end{quote}
By taking care of tasks that an individual might otherwise have to do, IoT devices can reduce cognitive loads on household members who have responsibility for these tasks and other members who want to ensure these tasks are completed in a timely fashion.

\subsubsection{Improving Peace of Mind}
Some of our participants reported that the convenience provided by IoT devices gave them peace of mind and eased specific worries.
According to PS379, who talked about devices for baby monitoring and security,
\begin{quote}
Having baby monitors and a smart lock really helps ease our worries, and as worries disappear, there is more room for good feelings.
\end{quote}
Peace of mind was also a commonly cited benefit among participants who
reported having IoT security systems, including security cameras and door locks.
PS8 talked about the feeling of safety provided by their IoT security cameras: 
\begin{quote}
The smart security cameras provide
us with peace of mind, and we feel safe to go out and do things together
knowing the house is being watched over.
\end{quote}
PS143 reported a similar effect from outdoor cameras and an alarm system easily accessed on a smartphone:
\begin{quote}
I have Ring floodlight cameras as well as a smart alarm system connected to my phone, which has given me and my spouse increased peace of mind regarding the security of our home.
\end{quote}
By allowing household members to monitor the state of their environment inside and outside the home, IoT devices made our participants feel more at ease.

\subsection{C1.~Facilitating Undesired Monitoring} 
\label{sec:monitoring}
While IoT devices facilitated many benefits, they also caused many conflicts. Some of our participants reported that they or other household members 
were worried about or had experienced surveillance by other household members 
via their IoT devices.
Devices our participants associated with unwanted
monitoring all enabled audio or video recording, including security cameras,
door bells/locks, and voice assistants. PS433 talked about how one of their
housemates became upset by having a Google Home in the house: 
\begin{quote}
My
housemate was very upset when we brought the Google Home in. He is concerned
with spying. We appeased him by turning off the microphone, but he has since
read that this is not effective.
\end{quote}
In another example, PI1 reported the potential for surveillance of household members without their knowledge:
\begin{quote} 
I was really shocked. I didn't know [the security camera] was working. I thought it was just put in as a design, you know, to threaten someone who's come [to rob the house]. But then when I found out it was tracking everything, I was really concerned.
\end{quote}
This led PI1 to address the roommate who had installed the cameras, but this household member ``asked me [PI1] not tell anybody.'' 
PI1 continued to describe how this monitoring could be of specific concern to roommates in relationships with others outside the house:
\begin{quote}
For other people in the house...they have some relation with other people outside the house. Probably someone here wants to know what's going on or when that person comes.
\end{quote}
Conflicts over the feeling of being monitored were also common among parents and children. As PI8 mentioned,
\begin{quote}
    [We have] about six security cameras set up in main areas mostly for security. But as my son has turned into a teenager, he thinks it's an invasion of privacy. So that's always an ongoing conflict even though that's not the intent of it. That's what he thinks. 
\end{quote}
PI10 also reported conflicts between parents and children over IoT monitoring, but from the opposite perspective:
\begin{quote}
My brothers had a party and it was really loud. So nobody heard that people had been ringing the doorbell. And my boyfriend actually was the first one to ring the doorbell for some reason. And you know when you ring the doorbell there's like a video recording, so my parents got a nice snapshot of my boyfriend bringing in like ten pizzas into the house. 
\end{quote}

Concerns about and instances of household surveillance using IoT devices were
 common in the survey responses as well. 40\% (204/508) of
respondents believed that Internet-connected products could allow people in
their household to monitor each others' behavior, and 9\% (46/508) reported
disagreements about the use of these products for monitoring. A further 15\%
(78/508) agreed that they had actually used these products to monitor others'
behaviors, and 9\% (45/508) agreed that someone else in their household had
used these products to monitor their own behavior
(Figure~\ref{fig:conflicts-likert}). 
Comparing across demographic groups, we found that respondents in households with four to six people were significantly more likely to report using IoT devices to monitor others' behavior than respondents in two-person households ($p < 0.01$). 

Other researchers have also found that being monitored in the household is often
perceived as a risk of IoT
devices~\cite{Wilson2017}, which could also lead to domestic
abuse~\cite{Bowles}. 
Given the increasing popularity of IoT
products, the prevalence of monitoring found in our survey means that many households are likely facing new
interpersonal conflict concerning actual or potential surveillance enabled by these
devices. 

\subsection{C2.~Provoking Differences in Knowledge or Preferences About IoT Devices} 
\label{sec:diff-prefs}
We found that a common cause of conflict between household members
involving IoT devices resulted from differing knowledge, opinions, or
preferences about these devices. Related work has
shown the effects of such differences on household power dynamics~\cite{Brush,Demeure,Mennicken2014,Mennicken,Zeng}; the rest
of this section offers more specifics and data about the prevalence of this cause of conflicts.

\subsubsection{Differing Interest in IoT Technology}
A few of our participants had disagreements among family members stemming from different levels of interest and perceived necessity of IoT technology. PS481 talked about disagreements over a smart TV:
\begin{quote}
My family and I have always had minor disagreements over our smart TV. My mother doesn't really like the features the TV has and complains about technology in general, saying it's over complicated.
\end{quote}
PS208 described a similar conflict around the expense and necessity of IoT devices:
\begin{quote}
My parents often argue about the cost of all these Internet-connected devices and if we really need them or not.
\end{quote}
In a few cases, arguments about IoT devices placed interests in home technologies directly at odds with perceived optimal conditions for others in the household. PS67 gave one such example of making a simple task more complicated unnecessarily:
\begin{quote}
My husband added smart bulbs and taped over all the light switches and switched us over to using Alexa to turn on and off the lights. I don't like it because there are times when my young children fall asleep and I want to turn off the lights silently instead of using my voice. My children don't like it because their pronunciation is not clear and Alexa cannot understand them sometimes when they want the lights on or off.
We have argued about it a couple of times but it has been made clear that his excitement for a smart home outweighs the desires of me and our two kids, so now I just deal with it and try to help my kids as much as possible.
\end{quote}
Prior work has examined how families with children attempt to repair communications breakdowns with Alexa voice assistants~\cite{beneteau2019communication}, due to pronunciation,  code-switching, or other linguistic factors. Our findings indicate that such communication breakdowns can lead to interpersonal conflict in addition to or instead of collaborative troubleshooting.
Overall, 14\% (71/508) of survey respondents reported disagreements between household members about whether they should have Internet-connected products in their homes, while 22\% (110/508) of survey respondents said that they were simply less interested in these products than others in their household (Figure~\ref{fig:conflicts-likert}). 

\subsubsection{Differing Concerns About Privacy and Security} Our participants
also had differing understandings and opinions of the privacy policies and security features of IoT devices. 
Some reported that different privacy and security attitudes
caused conflicts in their household. For instance, PS159 described 
disagreements about the privacy implications of an Amazon Echo Dot: 
\begin{quote}
My partner and I
had a disagreement over bringing in an Echo Dot into our household for privacy
reasons. I understood where he was coming from, but I thought the convenience
outweighed the possible concerns for privacy, as it is in a room we don't use
very often.
\end{quote}
PS403 reported a similar disagreement which resulted in them returning the device for privacy reasons:
\begin{quote}
I bought an Amazon Echo so I could play music with it. My wife was very nervous about it listening to our conversations. I decided to return it to make her more comfortable.
\end{quote}
PI2 indicated that disagreements about privacy and security issues often arise when different household members have different opinions about the value of new technology in and of itself:
\begin{quote}
Beforehand I was like `are you
insane...like is this 1984...we don't need this,' but he, like I said, he's a
tech guy. He's an early adopter. He likes to play with whatever the newest
thing is.
\end{quote}
PI2 also cited uncertainty about how to turn off the microphone on an Amazon Echo or how to use other privacy protection features:
\begin{quote}
When she [the Amazon Echo] says "I listen when I hear the wake word" does that mean she's off the rest of the time? Is that what that is? 
[My housemate] also is pretty into privacy so I'm sure whatever actions there were to scale back her monitoring or recording or whatever...I'm sure he chose them. But I don't know what they are.
\end{quote}
Overall, 22\% (112/508) of survey respondents disagreed with others
in their household about the privacy risks of Internet-connected products
(Figure~\ref{fig:conflicts-likert}). 

\subsection{C3.~Causing Tensions About Device Use, Sharing, and Technical Issues}
\label{sec:use-sharing}
About half of the participants who reported interpersonal conflicts due to their
IoT devices attributed this conflict to how these devices were being used
and shared in the household. 

\subsubsection{Disagreements About Sharing}
\label{sec:sharing}
The most common source of tension between household members was due to different family members wanting to use the same IoT device at the same time and disagreeing over who should have access. This was most prevalent among children and between children and parents. P141 described such a conflict:
\begin{quote}
It's basically just the sharing aspect as far as our children share certain devices sometimes and one child wants to use it a little longer than expected and that's where the disagreements come in. So now we are in the process of getting separate devices for our children.
\end{quote}
PI8 also said that simultaneous use of devices can affect the Internet connection more generally:
\begin{quote}
When [my son] is using all the devices it slows it down\dots [and when] I'm trying to work it slows down bandwidth\dots that's tough.
\end{quote}
A few of our survey respondents reported device sharing conflicts specifically involving IoT thermostats. These disagreements typically occurred between spouses and partners as in the following example from PS19:
\begin{quote}
My wife and I often disagree on how to program our Nest thermostat. She likes it to be 70 at night but I feel like that's too cold. Also, the Nest is using my wife's phone proximity to set its Eco Mode, so if I am home and she is not, then I have to take it off of Eco Mode and manually set the temperature.
\end{quote}
The multiple-choice survey responses also indicate issues with sharing, with 12\% (60/508) and 9\% (44/508) of respondents agreeing that who should have control of or access to Internet-connected devices, respectively, had caused disagreements in their households (Figure~\ref{fig:conflicts-likert}).

\subsubsection{Frustrations About Technical Issues}
Another common source of tension and arguments among household members
resulted from frustrations about technical aspects of IoT devices.
For example, PS170 described frustration over technical challenges of their IoT devices as a source of conflict with their partner: 
\begin{quote}
Either me or my partner sometimes get frustrated when we want to use a product and it isn't working correctly. Then we can take it out on each other.
\end{quote}
PS361 described a related situation where one individual's greater technical knowledge led to conflict between spouses sharing a device: 
\begin{quote}
My husband is not as tech savvy as me and gets irritated with me when I can get a device to do something he can't.
\end{quote}
In contrast, PS377 reported that their ability to troubleshoot voice assistants and IoT security cameras was appreciated by other household members but sometimes caused additional tension: 
\begin{quote}
My parents sometimes want things fixed that are beyond my control. We sometimes disagree about what products to purchase and how they would perform on our network.
\end{quote}
These individuals are not alone in dealing with conflicts related to technical issues of IoT devices.
20\% (102/508) and 15\% (74/508) of survey respondents agreed that these devices have caused tension in their households because they do not work as expected or are difficult to configure, respectively (Figure~\ref{fig:conflicts-likert}).

\subsubsection{Antagonistic Use of Devices}
\label{sec:antagonistic-use}
A few of our participants talked about how their IoT devices were used to disrupt and annoy other household members in new arguments and pre-existing conflicts. 15\% (75/508) of the survey respondents agreed that these devices were contributing to tensions in their households unrelated to the technology itself (Figure~\ref{fig:conflicts-likert}). 
For example, PI11 reported the involvement of an Amazon Echo in unrelated arguments: 
\begin{quote}
    Any time that we try to have a conversation about not using our phones or anything like that, the biggest thing is that mostly my fiance, he turns on Alexa and asks her to play a song and at a really high volume so he can't hear me talk anymore\dots Sometimes it's really frustrating and sometimes it actually diffuses us because he'll play music.
\end{quote}
A parent, PS68, described how their Amazon Echo became a source of fights for their children: 
\begin{quote}
Our young children `fight' over talking to Alexa. They use Alexa to play songs and will cancel the other one's music, or ask her to repeat them and use her to insult one another.
\end{quote}
Another type of IoT device misuse was related to children ordering products online without their parents' permission. PI4 reported this behavior when talking about their experience with Amazon Echo and how their son used it without their knowledge: 
\begin{quote}
My youngest son has ordered toys or put hundreds of dollars of toys in our Amazon cart and we just caught it at the last second.
\end{quote}
These examples indicate that conflict connected to consumer IoT devices can originate both from the devices themselves as well as from the use of the devices to perpetuate or escalate other interpersonal tensions.

\subsection{Conflict Mediation}
\label{sec:mediation}

\begin{figure}[t]
    \centering
    \includegraphics[width=0.75\textwidth]{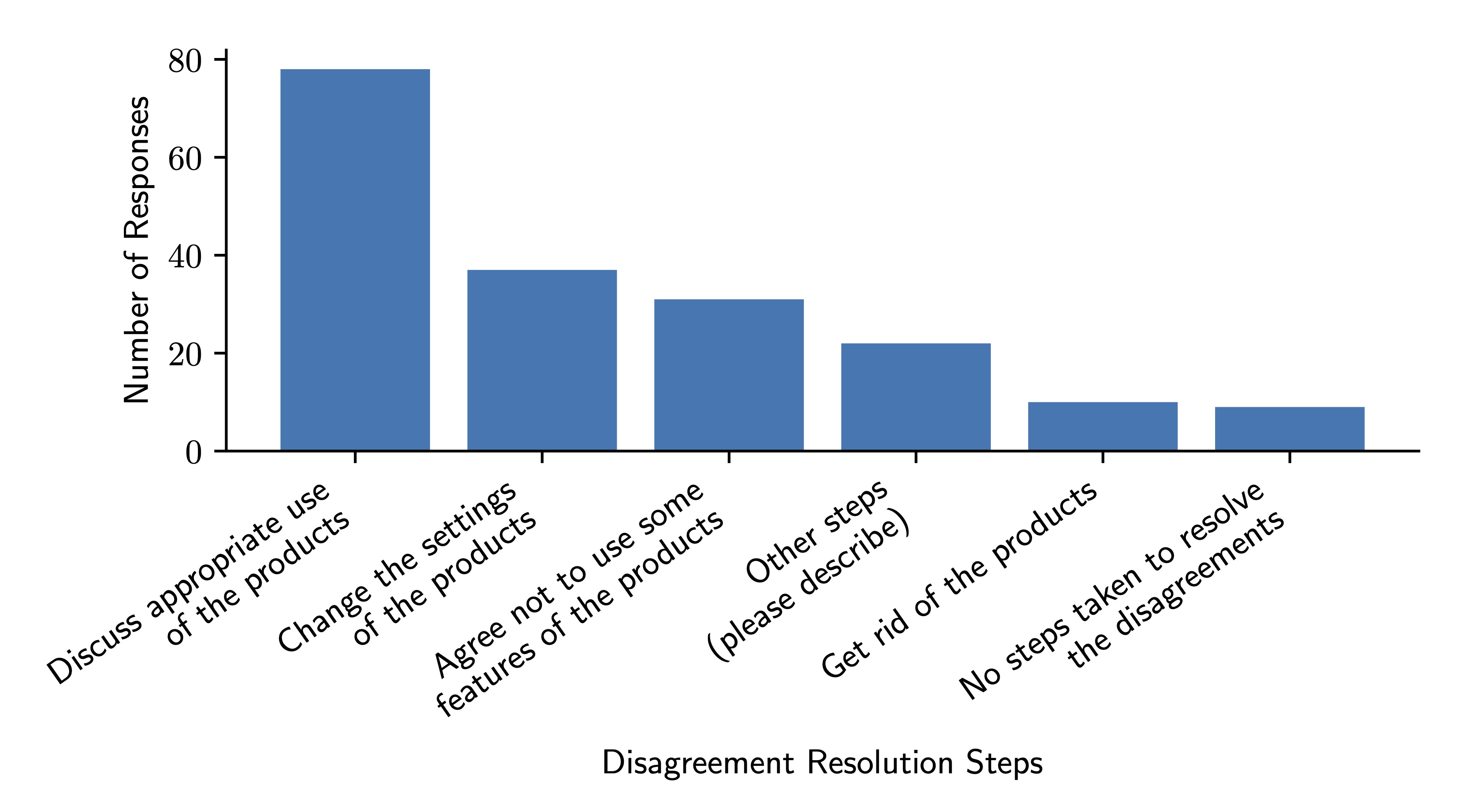}
    \caption{Prevalence of conflict mediation strategies among survey respondents who reported disagreements with others in their households regarding IoT devices. }
    \label{fig:disagreement-resolution}
\end{figure}

Our participants reported several different methods for mediating conflicts
involving IoT technologies. Figure~\ref{fig:disagreement-resolution}
presents the frequency of mediation strategies used by survey respondents who also reported disagreements between household members caused by Internet-connected products. Discussing appropriate use was the most common strategy, followed by settings changes and agreeing not to use certain features of the products. 
For example, PI1 described a conversation about the placement of security cameras to keep household members from feeling uncomfortable:
\begin{quote}
When [my roommate] was setting up the cameras, he proposed to have one camera downstairs like around the entrance. But I said, ``No, this is not polite at all to have the camera inside, because it would be like tracking someone's motion, or sometimes you might be dressed in a certain way around the house.'' So I said, ``I think we are very close to each other, and we should not do that in the house.'' So we don't have...as much as I know...there's [no camera] in the house.
\end{quote}
Other participants gave other examples of these strategies, including discussing communication issues exacerbated by IoT devices (PI11) and agreeing on schedules for device and bandwidth sharing (PI8).
Strategies encapsulated in the ``other steps'' category (Figure~\ref{fig:disagreement-resolution}) include placing the device in a little-used room (PS233), increasing household Internet speed (PS494), and setting consequences if children misused the devices (PS492).  
\section{Discussion and Future Work} \label{sec:discussion}

In this section, we explore the implications of our findings concerning 
interpersonal benefits and conflicts arising from
consumer IoT.  We first explore possible future research
avenues stemming from the benefits that this study has uncovered. We then
explore how the conflicts might be mitigated through future research.

\subsection{Amplifying benefits}

Our findings suggest that IoT devices can benefit interpersonal relationships
by empowering individuals, 
facilitating certain management tasks, and strengthening interpersonal bonds. 
Each of these findings offers immediate
implications and opportunities for future work.

\subsubsection{Design for strengthened interpersonal connections.} Previous work has
explored the extent to which home automation provides a sense of
emotional comfort, as well as how consumer IoT devices can make people feel
safer and more secure. Our work builds on these previous results, with nearly half of respondents indicating
that IoT devices in the home improved interpersonal relationships through
shared experiences, improved communication, and playfulness. 
Respondents described positive interpersonal experiences involving 
IoT devices using terms like ``bonding'' (PS97, PS73), ``laugh'' (PI7), ``joy'' (PI2),
``connected'' (PS381), and ``communicating'' (PS21). Devices that 
reduced the technological complexity or time overhead required for
users to engage with others
or encouraged playful feature exploration were often involved in these positive experiences.
Our study did not, however, dissect which specific design elements
are most likely to lead to these positive
outcomes or the specific devices that were most likely to cause these
positive benefits. Future research could disambiguate these elements and
explore how new IoT devices could further enhance the positive benefits we observed, such as by making it easier for users to have shared experiences or by
directly encouraging playfulness through interfaces \cite{whereaboutsclock} or nudging. 

\subsubsection{Understanding how design affects empowerment.} One of our 
more surprising findings was that individuals, especially older adults and those with accessibility needs, 
experienced a sense of increased
independence and empowerment (Section~\ref{sec:independence}). 
For example, two participants (PI4, PS445) reported that IoT interfaces were particularly
helpful for special needs children in their households who could independently find media 
content and other information through the devices.
Although this finding emerged as a theme in this
study, the benefits of empowerment likely depend on context, as well as the
nature of the specific devices that are deployed in a shared household
setting. This finding is particularly interesting because it runs counter to
existing work that explores the more pernicious effects of shared
IoT devices (e.g., intimate partner violence \cite{Geeng, Freed})
and our own result that devices can provide unbalanced utility for 
different household members (Section~\ref{sec:diff-prefs}). 
Future studies could further explore the circumstances under which devices might
empower or disempower an individual in a shared household setting.
One potential hypothesis to explore is the relationship
between an individual's {\em autonomy} and their sense of empowerment. For
example, it is possible that an individual may feel more empowered if they have
some control over how a particular device is deployed and used as
well as how it collects data about them and others
in the household. This hypothesis is supported by the multiple 
participants who expressed dissatisfaction with the IoT devices in their household
related to a perceived loss of control or limited understanding of the devices (Sections~\ref{sec:monitoring}-\ref{sec:use-sharing}).

\subsubsection{Technology design for easier home management and automation.} Our
findings revealed that consumer IoT technology can provide benefits by making it
easier for household members to coordinate management tasks and by increasing
free time to spend with one another (Section~\ref{sec:easing-tasks}). 
This suggests that designing devices
with household management in mind could not only ease the 
home management responsibilities of individuals but
correspondingly benefit interpersonal relationships
of users sharing home management tasks. Of particular interest is the result that shared
management interfaces can reduce arguments about various management tasks
(e.g., locking doors). This suggests that, if certain technologies are
deployed more broadly, these effects could be more pervasive across household
chores, from cleaning to groceries. Of course, such pervasive deployment also
carries associated privacy risks, and thus it is important that future
research consider these benefits in light of potential conflicts, as we
discuss in the next section.

\subsection{Mitigating conflicts}

This study revealed three themes concerning interpersonal conflicts resulting
from shared IoT devices: the potential for surveillance and
mistrust, unease as a result of differences in knowledge or preferences, and
tensions surrounding shared use of devices. We now explore various
opportunities for future work concerning each of these findings.

\subsubsection{Mitigating surveillance risk.} Consumer IoT devices 
unilaterally increase opportunities for surveillance---not only by third
parties, but also by other household members. This concern emerged as a
significant source of conflict in this study (Section~\ref{sec:monitoring}), which echoes and amplifies a large
body of previous work on IoT privacy, tracking, and intimate partner violence~\cite{Geeng, Freed}.
If this surveillance risk is not
mitigated, shared IoT devices could further exacerbate existing power imbalances in
domestic settings---particularly in situations where users may have limited
autonomy. For example, a roommate may have limited autonomy over what devices another
household member deploys in the house, creating a situation of unwanted
or unknown surveillance such as that described by participant PI1.
A child or teenager may have limited autonomy over audio or visual recording devices
installed by their parent or guardian as described by participants PI8 and PI10.
A victim of intimate partner violence
may not even be aware of the deployment of certain technology, let alone have
the capability to control its deployment and use. Such settings may result in
IoT devices either amplifying a lack of trust or a power imbalance that already
exists or introducing a new one. Future work
must focus not only on understanding these risks but also on allowing
users to mitigate them whenever possible. Furthermore, mitigation technology
should not be cumbersome or difficult to use. Recent work from Chen et
al.~\cite{chen2020wearable} on wearable microphone jamming is one such approach
for preventing IoT devices from recording audio. More work is needed in this area to provide users with
usable technologies to mitigate in-home surveillance.

\subsubsection{Improving user understanding of device function.} Many household
conflicts arise because different members of the household have different
understandings of a device's function and may thus reach entirely different
conclusions about the benefits and risks of a particular device (Section~\ref{sec:diff-prefs}).
This is supported by our findings, as well as by prior work focusing specifically on IoT voice assistants~\cite{garg2019exploring}.
Ultimately,
even with the same set of facts, different household members may view
associated benefits and risks differently, merely as a result of different
values or priorities. 
Nevertheless, our findings suggest that some
conflicts could be mitigated if users at least had a common understanding of a
device's function, as well as a basic understanding of how to use, reset, and
even disable the device if desired. To draw an analogy to the physical world,
different household members may have different views on the appropriate
thermostat setting, whether to keep the blinds open or closed, or whether to
turn off the lights when leaving a particular room---such conflicts are
inherent, but can be surfaced more directly because all participants know how
to operate devices such as blinds and light switches.  Similarly, IoT
devices could provide ``quick start'' guides to any user who installs an
application on their mobile device to allow all household members to be
apprised with the same information about basic function and operations.

A related approach could be to make interaction with IoT devices more
tangible.  For example, webcams can be equipped with physical covers,
and most voice assistants have mute buttons to stop continuous recording.
Related research in HCI
is already exploring how similar tangible interfaces for consumer IoT devices can make managing privacy with these devices more intuitive~\cite{ahmadtangibleCSCW20}. Future work could also explore how these tangible interfaces can be designed to provide ``useful intelligibility''~\cite{mennicken2014today} specifically to mitigate
conflict in interpersonal relationships. 
We expect that several of the conflicts reported in Section~\ref{sec:monitoring} could have been avoided by improved
notifications indicating to all household occupants when certain monitoring features were active. The exact details of these interfaces would vary by device, but our findings show that they must be accessible and intelligible to all household members, including those not involved in device setup or management.

\subsubsection{Designing for conflict mediation.} 
We observed conflicts concerning the use of shared devices and resources, from
thermostats to Internet connectivity (Section~\ref{sec:sharing}). Past work has demonstrated that
making information about resource usage or actions more transparent can help
reduce conflicts~\cite{chetty2011my}. Future research could extend this past
work into the home IoT setting to better understand whether and how
exposing information about device usage and interaction could help mitigate
certain sharing conflicts.
This research could also reference prior work seeking to provide transparency 
for mitigating privacy threats in IoT systems~\cite{seymour2020}.

Our study found that individuals sometimes use IoT devices to antagonize other members of their
household, such as by using a voice assistant to play a song at high volume (Section~\ref{sec:antagonistic-use}).
These anecdotes highlight the difference between conflicts caused by devices themselves and unrelated conflicts 
exacerbated by device use. While household conflict 
pre-dates consumer IoT, future research could explore
interfaces or nudges that discourage the use of these devices to escalate antagonistic behavior towards other household members. 
This work could draw from prior studies of IoT device use unanticipated by designers~\cite{Morris}.

\subsection{Designing for diversity}

Households can have many types of relationships, including parents
and children of varying levels of independence, intimate partners with
individual insecurities and task responsibilities, inter-generational families
with different levels of technological familiarity, and many other unique
situations. Our results provide further evidence that many IoT devices do not
provide settings options with enough flexibility to account for this
variety of relationships among household members. In particular, our results
suggest that parent/teenager, roommate, and older adult/caregiver
relationships are especially poorly-served by the default ``adult partners
with or without young children'' model assumed by many device manufacturers.
In the case of parents and teenagers, IoT devices can cause conflicts when
there is unintended surveillance of teenagers who are in a transitional stage
of independence (Section~\ref{sec:monitoring}). 
When device features do not allow for more complex sharing
situations, users must revert to social resolution techniques to negotiate
device use, such as agreeing not use some features or engaging in long-term
discussions about appropriate interactions with a device (Section~\ref{sec:mediation}). 

Our findings support existing evidence~\cite{Kraemer} that IoT device users
employ a variety of social and technical approaches to address potential and
actual interpersonal conflict arising from these technologies.  One potential
path forward is to offer additional default settings that cater to common household
relationships beyond the nuclear family. For instance, the initial setup for a
voice assistant could involve choosing between ``roommates,'' ``frequent
visitors,'' ``caregiver,'' ``nuclear family,'' or other such defaults, allowing users in
those situations to select these options instead of creating and managing
separate accounts for every user---a task that often seems
overwhelming due to the technological familiarity required for configuration
and the ongoing attention required to use the correct account when many users
share devices fluidly. Designing these default settings would force device
manufacturers to consider whether their devices are able to
gracefully handle a diversity of household scenarios or what additional
functionalities might be required. 
This approach may also inspire further research into what
default settings would best cater to specific household
situations.
As long as these defaults are
well-explained during the setup process and provide some flexibility for
unique circumstances, they could reduce the prevalence of interpersonal
conflicts involving IoT devices.

\section{Limitations}
\label{sec:limitations}

This study has the following limitations, mostly due to the qualitative nature of the interviews, potentially sensitive topic of the research, and representativeness of the participants.

Some interview participants may not have felt comfortable sharing details of their interpersonal relationships with researchers. However, 
the follow-up survey provided a more anonymous setting for participants, allowing us to uncover additional benefits and harms to interpersonal relationships.
Some participants may also have become used to their IoT devices over time and been unable to remember their interpersonal impacts. However, this possibility emphasizes the importance of this research, suggesting that the impact of IoT devices on household relationships may have an even broader scope than we report. 

Self-reported demographics indicate that, while diverse, our interview and survey participants were still non-representative in ways that may bias our results. For example, our participants were skewed toward a younger demographic. We chose not to compare our findings across age groups to avoid conflating factors, as our participants often lived households with older or younger members. However, a 2017 survey~\cite{Liu} did observe that 46\% of IoT device owners were 26-35 years old, similar to the age range of our participants. 

Additional demographic characteristics that we did not collect, such as participant race and elements of socioeconomic status other than income, have also been shown to correlate with technology use by parents and children. Garg, et al.~\cite{Garg2019you} reported these effects for IoT speakers and smartphones, and it follows that they would carry over to other IoT devices as well. Shin et al.~\cite{Shin} point out that the characterization of ``the home'' in human-computer interaction literature remains narrow and typically does not include alternative domestic configurations, such as collective homes, that are also not represented in this work.
These limitations emphasize the exploratory nature of our findings and the need for future research focusing on specific interpersonal effects of IoT technologies in targeted populations. 

Our observed prevalence of interpersonal benefits over conflicts may also be due to a participant selection bias. Participants who have decided to purchase and continue using IoT devices may have disproportionately positive sentiments towards these technologies~\cite{Apthorpe-discovering}. Future research is needed to understand the experiences of users who choose to avoid or discontinue use of IoT products. Users responsible for the setup and maintenance of the IoT devices in their homes may also have been more likely to respond to our recruitment advertisements. Future research could explicitly recruit participants who live with IoT devices but who were not involved in purchasing or deployment decisions.
\section{Conclusion} 
We conducted semi-structured interviews of 13 participants and a followup survey with 508 respondents to understand the impact of consumer IoT devices on interpersonal relationships in multi-occupant households. We identify and categorize the most pervasive positive and negative impacts of consumer IoT devices on participants' relationships with other household members. 

On the positive side, we find that IoT devices strengthen interpersonal connections through bonding over shared experiences, simplify remote
communication, inspire playfulness, support independence of individuals with special needs, ease household management, improve peace of mind, and increase free time to spend with household members.
On the negative side, we find that IoT devices facilitate surveillance and cause mistrust due to potential or actual undesired monitoring and a lack of data collection transparency, provoke differences in knowledge or preferences about the functionality, benefits, risks, privacy, or
security of the devices, and cause tensions about device use, sharing, and technical issues that arise during day-to-day operation.

These findings suggest design improvements that would amplify the interpersonal benefits of consumer IoT devices, prevent or mitigate many of the reported conflicts, and support greater diversity of household relationships. For example, devices should more readily support sharing arrangements for multi-generational families and non-familial roommates. Devices should also provide clearer descriptions of data collection behavior to limit conflicts arising from different views of surveillance potential.
This paper also informs future research, motivating studies of users who have chosen \textit{not} to incorporate IoT devices into their households and closer examinations of IoT devices supporting independence and empowerment.


\begin{acks}
We thank our study participants. This work was funded by NSF Award CNS-1953740.
\end{acks}

\Urlmuskip=0mu plus 1mu\relax
\def\UrlBreaks{\do\/\do-}
\bibliographystyle{ACM-Reference-Format}
\bibliography{IoT-Relationships.bib}


\end{document}